\documentclass[]{IEEEtran}
\usepackage{graphicx}
\usepackage{cite}
\usepackage[version=3]{mhchem}
\usepackage{amssymb}
\usepackage{color}
\usepackage{algorithm}
\usepackage{algpseudocode}
\usepackage{amsmath, multicol}

\DeclareMathOperator{\erfc}{erfc}
\newcommand{\Ugtrless}{%
	\mathrel{\kern0pt\mathop{\gtrless}\limits^{1}_{0}}%
}
\usepackage{textcomp}
\begin{document}
	\title{Design and Evaluation of a Receiver for Wired Nano-Communication Networks}
	\author{\IEEEauthorblockA{Oussama Abderrahmane Dambri\textit{, Student Member, IEEE}  and Soumaya Cherkaoui\textit{, Senior Member, IEEE}}
	}
	\maketitle
	\begin{abstract}
		In this paper, we propose a bio-inspired receiver, which detects the electrons transmitted through a nanowire, then, it converts the detected information into a blue light using bioluminescence. Using light allows the designed receiver to also act as a relay for the nearest gateway (photo-detector). We simulate the construction of the nanowire, present its electrical characteristics and calculate its maximum throughput for a better design of the receiver. The designed receiver contains two parts, a part that detects the transmitted electrons, which we model by using an equivalent circuit, and a part that converts the detected electrons into a blue light. We derive the analytical expressions of the equivalent circuit's components, and we calculate the emitted photons for each electrical pulse detected. We also propose modulation techniques that guaranty an effective decoding of the information. We send a binary message and we follow the electron detection process of the proposed receiver until light emission and we calculate the Bit Error Rate (BER) to evaluate the performance of the designed receiver. The results of this study show that the designed receiver can accurately detect the electrons sent through a conductive nanowire in wired nano-communication networks, and that it can also act as a relay for the nearest gateway.
	\end{abstract}
	
	\section{Introduction}
	\IEEEPARstart{N}{anotechnology} has become a key area of research in multidisciplinary fields, and its rapid and impressive advance has led to new applications in biomedical and military industries. It turned out that nanomachines need a cooperative behavior to overcome their limited-processing capacities in order to achieve a common objective. Exploiting the potential advantages of nano-communication networks between the nanomachines drove the researchers to come up with new solutions to create systems that communicate at nanoscale. Three methods are proposed in literature, wireless electromagnetic method using THz band \cite{akyildiz_internet_2010, piro_terahertz_2015, doro_timing_2015, yao_tab-mac:_2016}, bio-inspired wireless molecular communication using molecules \cite{farsad_comprehensive_2016, pierobon_physical_2010, kim_symbol_2014, dambri_MIMO_2019, koo_molecular_2016, noel_improving_2014, dambri_performance_2019, tepekule_isi_2015, chang_adaptive_2018} and wired nano-communication method using polymers \cite{dambri_design_2019}.

	However, on the one hand, THz band suffers from a very high path loss, mainly caused by water molecules absorption, which complicates its use in medical applications inside the human body \cite{piro_terahertz_2015, doro_timing_2015, yao_tab-mac:_2016}. On the other hand, molecular communication biocompatibility makes it very promising for medical applications at nanoscale, such as monitoring, diagnosis and local drug delivery \cite{farsad_comprehensive_2016}. Nevertheless, the achievable throughput of molecular communication systems is very limited and the delay is very high due to the random distribution of molecules in the medium \cite{pierobon_physical_2010, kim_symbol_2014, dambri_MIMO_2019, koo_molecular_2016}. Moreover, molecular communication systems suffer from intersymbol interference, despite the proposed techniques in literature to reduce it \cite{noel_improving_2014, dambri_performance_2019, tepekule_isi_2015, chang_adaptive_2018, dambri_enhancing_2018}.
	
	Wired nano-communication is a new proposed method that uses the self-assembly ability of some polymers inside living cells to build an electrically conductive nanowire \cite{dambri_design_2019}. This method has the biocompatibility of molecular communication, but also, it has a very high achievable throughput since fast electrons are the carriers of information. One of the main challenges of the wired nano-communication system is detecting the electrons at the receiver without losing its biocompatibility. The proposed receivers in the literature for molecular communication are designed to observe or absorb molecules in the medium \cite{noel_active_2016, Bao_Channel_2019, Kilinc_Receiver_2013, Qian_Molecular_2019}, and thus, they cannot be used to detect electrons for wired nano-communication.

	In our recent works \cite{dambri_design_2019, dambri_Physical_2020}, we proposed the idea of a receiver for a wired nano-communication system without studying or evaluating it. In this paper, we present the detailed system model of the proposed receiver, and we evaluate its performance. The proposed receiver contains a Smooth Endoplasmic Reticulum (SER), which plays the role of Ca$^{2+}$ storage inside living cells \cite{koch_endoplasmic_1990}. The receiver also contains a recyclable concentration of a photo-protein \textit{Aequorin}, which can be found in the jellyfish \textit{Aequorea Victoria} that lives in North America and the Pacific Ocean \cite{shimomura_short_1995}. The electron detection at nanoscale is extremely difficult due to the quantum trade-off between information and uncertainty. Therefore, the designed receiver uses the transmitted electrons to stimulate SER and trigger a chemical reaction that generates a bioluminescent light by using the photo-protein Aequorin, as shown in Fig. 1. In order to study SER stimulation with the transmitted electrons, we simulated the conductive nanowire used in wired nano-communication systems, we presented its electrical characteristics and we calculated its maximum throughput. Bioluminescence is a chemical reaction used by living organisms to generate light with enzymes and photo-proteins. Converting the transmitted electron pulses into light pulses at the receiver makes the extraction of information easier and can create a link between nanoscale and macroscale communication systems.

	The rest of the paper is organized as follows. In section II, we present and explain the 2 algorithms used in our framework to simulate the nanowire’s assembly in a stochastic 3D system. We also present the electrical characteristics of the nanowire and we calculate its maximum throughput and the error probability. Section III highlights an in-depth description of the designed receiver, by clarifying the SER role in electron detection, and by explaining the triggered chemical reaction that emits a blue light. In section IV, we model the designed receiver by using an equivalent circuit, where SER and its membrane represent a capacitor and the sum of Ca+2 channels represent a resistance linked in parallel. We derive the analytical expression of the circuit components and we calculate the emitted photons for each electron pulse detected. We also calculate the Bit Error Rate (BER) of the designed receiver. In section V, we propose modulation techniques that guaranty an effective decoding of the detected information at the designed receiver. Section VI discusses the numerical results of this study, by following the detection process of a random binary message at the receiver until its conversion into a blue light. It also presents the results of BER to evaluate the performance of the designed receiver. We conclude the paper in section VII. 
	
	\section{Simulation Framework}
	
	GlowScript VPython is an easy and efficient way to create 3D environments for real time simulations. Our framework applies GlowScript as a browser-based implementation that runs a VPython program by using RapydScript, which is a Python-to-JavaScript compiler. In this study, we simulate the self-assembly of a polymer called "actin" to create a nanowire in real time, by using sphere molecules that diffuse randomly in a 3D cube as shown in Fig. 2. By using this framework, we can follow the position of the last assembled molecule in real time to study the speed of the nanowire's construction as the graph in Fig. 3 exhibits. In the next subsections, we present and explain the two algorithms that we used in our framework. 
	
	\subsection{Collision Between Actin Molecules}

	The algorithm 1 is a combination of two strategies, Periodic Interference Test (PIT) and a Predicted Instant of Collision (PIC). The (PIT) consists of checking to detect if any collision has occurred between two molecules, at every frame in the simulation. It tests if the molecules are approaching each other, and whether they interfere spatially. When a collision is detected, the algorithm computes the new velocities after the shock using the momentum and energy conservation equations; and then attributes the new velocity to the moving molecules after the collision. If the simulation detects no collision in that frame, it continues to the next one. The PIC, on the other hand, pre-calculates the exact time of collisions before spatial interactions between the molecules. The difference between PIT and PIC approaches is that the later perfectly models the collisions, at the cost of some extra computation, compared to the PIT approach, which is a lazy collision detection strategy. In our algorithm 1, we used a combination of the two approaches to perfectly detect and model the collisions in real time.

	\begin{figure}
		\centering
		\includegraphics[width=\linewidth]{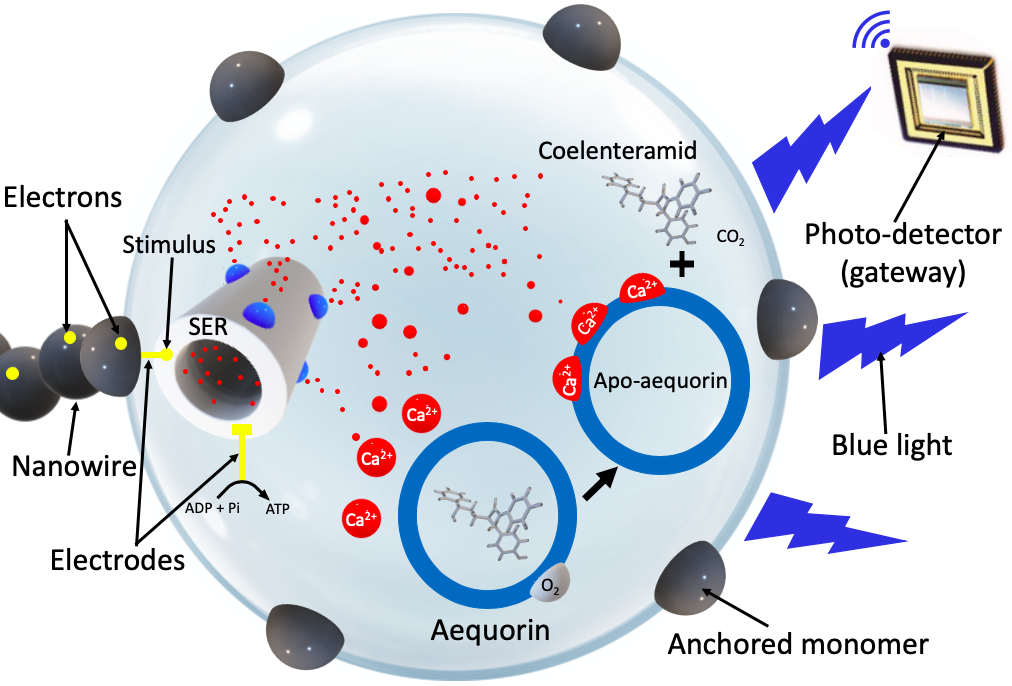}
		\caption{The designed receiver for wired nano-communication networks.}
		\label{fig:1}
	\end{figure}
	
	\begin{algorithm}
		\caption{Periodic Interference Test}
		\begin{algorithmic}[1]
			\Procedure{PITC( Molecule A, Molecule B ): }{}
			\State \text{At every frame instant t: }
			\State \text{Test if the two molecules are approaching each other.}
			\If {$\textit{YES}$}
			
			\State Test if they overlap (interfere) spatially.
			\If {$\textit{YES}$} 
			\State Collision detected.
			\State Compute New Velocities.
			
			\State \text{T:= time of collision between A and B.}
			\If {T $<$ next frame instant:} 
			\State \text{Move frame to intermediate instant T.}
			\State \text{Calculate velocities after collision.} 
			\EndIf
			\Else 
			\State \text{No collision detected. }
			\EndIf
			\State \text{Goto next frame instant.  }
			
			\Else:
			\State \text{Goto next frame instant. }
			
			\EndIf
			\EndProcedure
		\end{algorithmic}
	\end{algorithm}

	\subsection{Nanowire Formation}
	
	If a random molecule in the environment strikes the last molecule of the constructed nanowire, it will attach to it increasing the length of the nanowire. Several conditions are necessary for the molecules to stick in the nanowire, namely:
	
	\begin{figure}
		\centering
		\includegraphics[width=.96\linewidth]{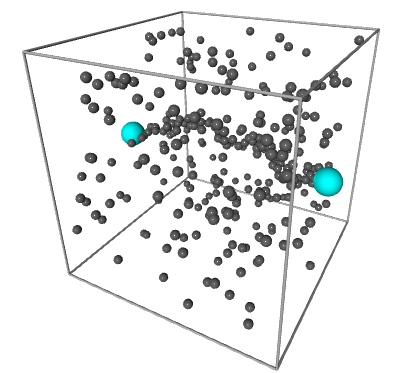}
		\caption{Actin-based nanowire formation represented by small black spheres, which link the transmitter to the receiver (cyan spheres).}
		\label{fig:7.1}
	\end{figure}
	
	\begin{itemize}
		\item One of the molecules having a collision is already sticking in the nanowire.
		\item The center of the striking molecule lies within a specified angle range from the center of the last attached molecule.
		\item The center of the next molecule being attached should be farther away from the transmitter and closer to the receiver, so that the nanowire is moving towards the receiver and not towards any random path.
	\end{itemize}
	
	\begin{algorithm}
		\caption{Nanowire Formation}
		\begin{algorithmic}[1]
			\State molecule\_array = [{Transmitter}]
			\If {collision.detected = True}
			\If{collision occurs with last(molecule\_array) \&\& last(molecule\_array) != Receiver}
			\State M $\leftarrow$ molecule hitting the wire  
			\If {Magnetic Field = 0}
			\State Randomly attach M to last(molecule\_array)
			\State molecule\_array.append(M)
			\EndIf
			\If {Magnetic Field != 0}
			\If {M.center.X $>$ max(last(molecule\_array) .position)}
			\State Z = function(magnetic field)
			\If {M.center.Z $<$ max(Z) }
			\State  M.momentum = 0      \% Stick the two molecules together
			\State molecule\_array.append(M)
			\EndIf
			\EndIf
			\EndIf
			\EndIf
			\EndIf
		\end{algorithmic}
	\end{algorithm}
	
	To make the molecules stick, we give their momentum a zero value, and we do not update their position with time. The algorithm 2 also regulates the direction of the nanowire formation. The direction is controlled by the magnetic field. If the magnetic field is zero, then there is no guide for the direction, and the nanowire forms randomly. As the magnetic field increases, the nanowire gets more and more aligned to the straight line joining the transmitter and the receiver, as confirmed by the experimental results in 	\cite{kaur_low-intensity_2010}.

	\begin{figure}
		\centering
		\includegraphics[width=\linewidth]{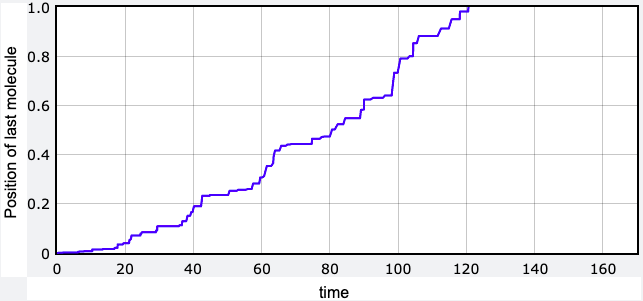}
		\caption{Position of the last assembled molecule as a function of time representing the speed of the nanowire formation.}
		\label{fig:8.1}
	\end{figure}
	
	\begin{figure}
		\centering
		\includegraphics[width=\linewidth]{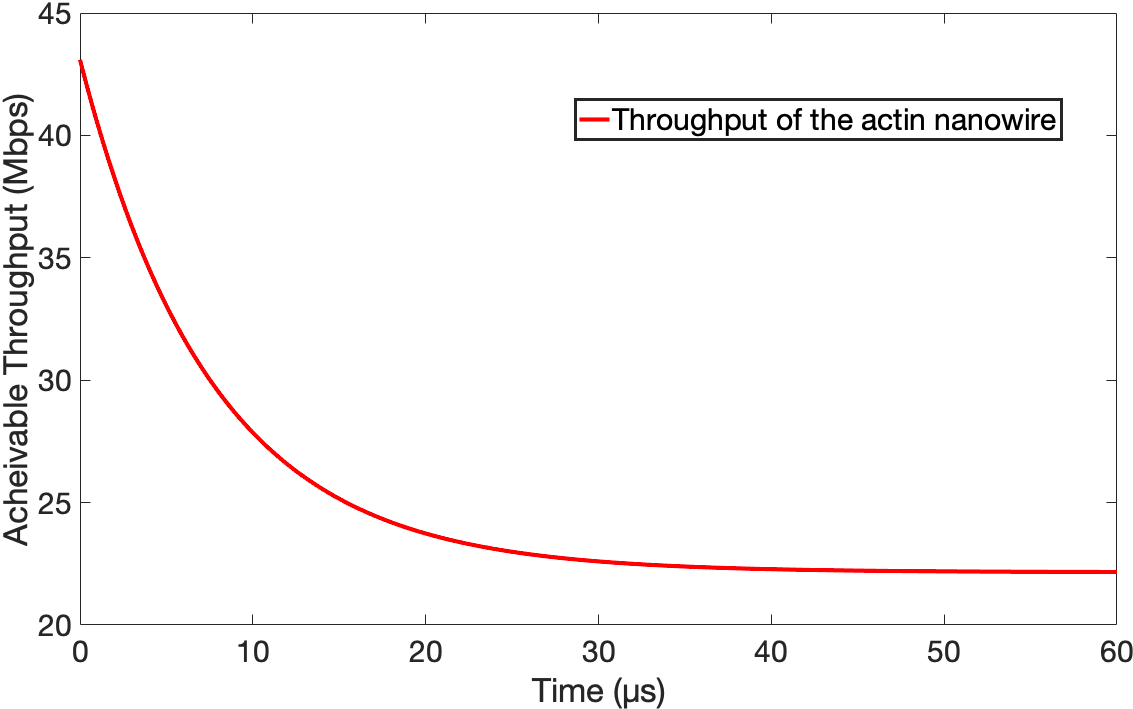}
		\caption{Maximum throughput approximation of the actin nanowire.}
		\label{fig:10.1}
	\end{figure}
	
	\subsection{Channel's Electrical Characteristics}
	
	The authors in \cite{hunley_multi-scale_2018} and \cite{tuszynski_ionic_2004} studied the electrical impulses and ionic waves propagating along actin filaments in both intracellular and in vitro conditions, by modeling the actin filament as an RLC equivalent circuit. The effective resistance, inductance and capacitance for a 1$\mu m$ actin filament are \cite{tuszynski_ionic_2004}:
	
	\begin{flushleft} 
		\centering
		$C_{eq}$ = 0.02  $\times$ 10$^{-12}$ F
		
		$L_{eq}$= 340  $\times$ 10$^{-12}$  H
		
		$R_{eq}$= 1.2 $\times$  10$^{9} \Omega$
	\end{flushleft} 
	
	The Fig. 5 represents actin nanowire's electrical characteristics simulated in our recent work \cite{dambri_toward_2019}, in terms of attenuation, phase and delay as a function of the frequency and the distance between the transmitter and the receiver. We can see that the attenuation of actin filaments is very high because of its resistance. The experimental study in \cite{hunley_multi-scale_2018} showed that despite the actin filaments attenuation, the velocity of the charges passing through it can reach 30,000$\mu m$/$s$ and then slows down rapidly in the first 60 $\mu s$. Moreover, the authors in \cite{patolsky_actin-based_2004} proved experimentally that adding gold nanoparticles to the actin monomers drastically decreases its attenuation.
	
	\begin{figure*}[!htb]
		\minipage{0.32\textwidth}
		\includegraphics[width=\columnwidth]{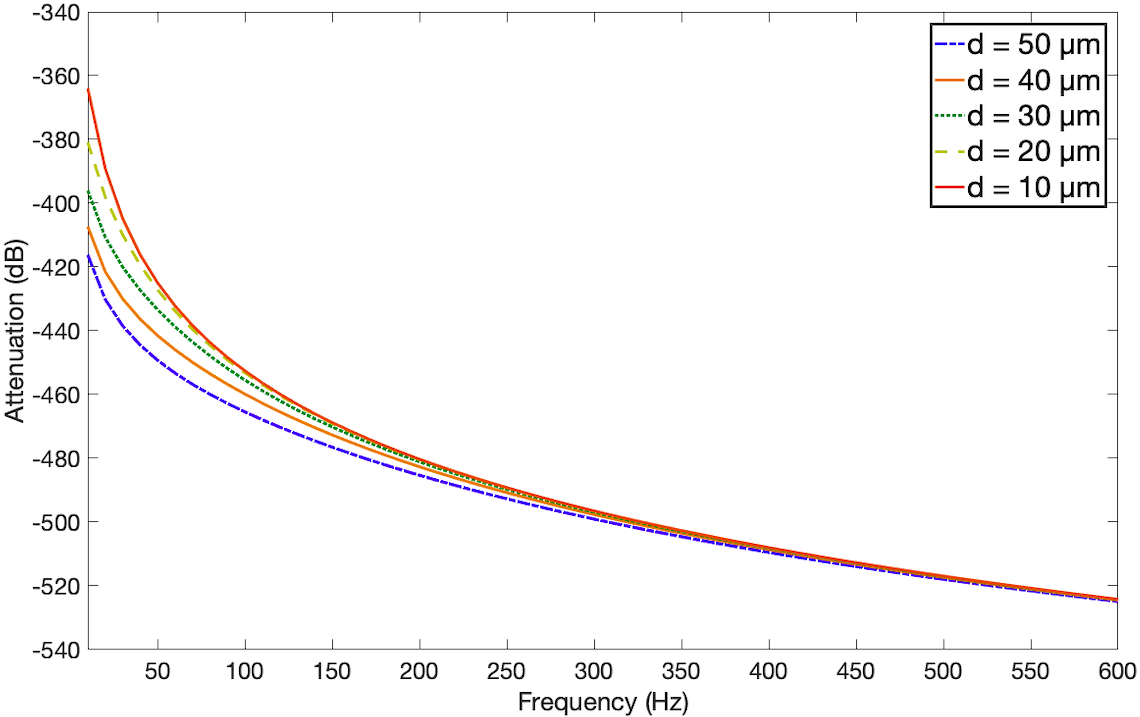}
		\endminipage\hfill
		\minipage{0.32\textwidth}
		\includegraphics[width=\linewidth]{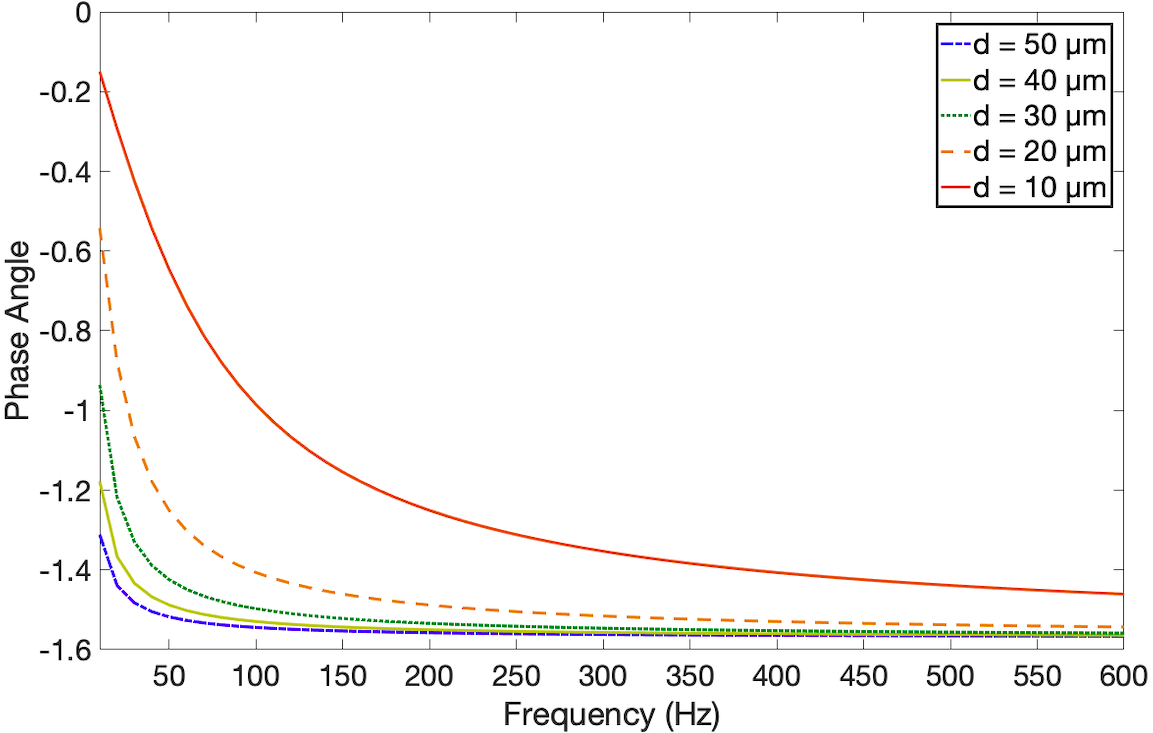}
		\endminipage\hfill
		\minipage{0.32\textwidth}
		\includegraphics[width=\linewidth]{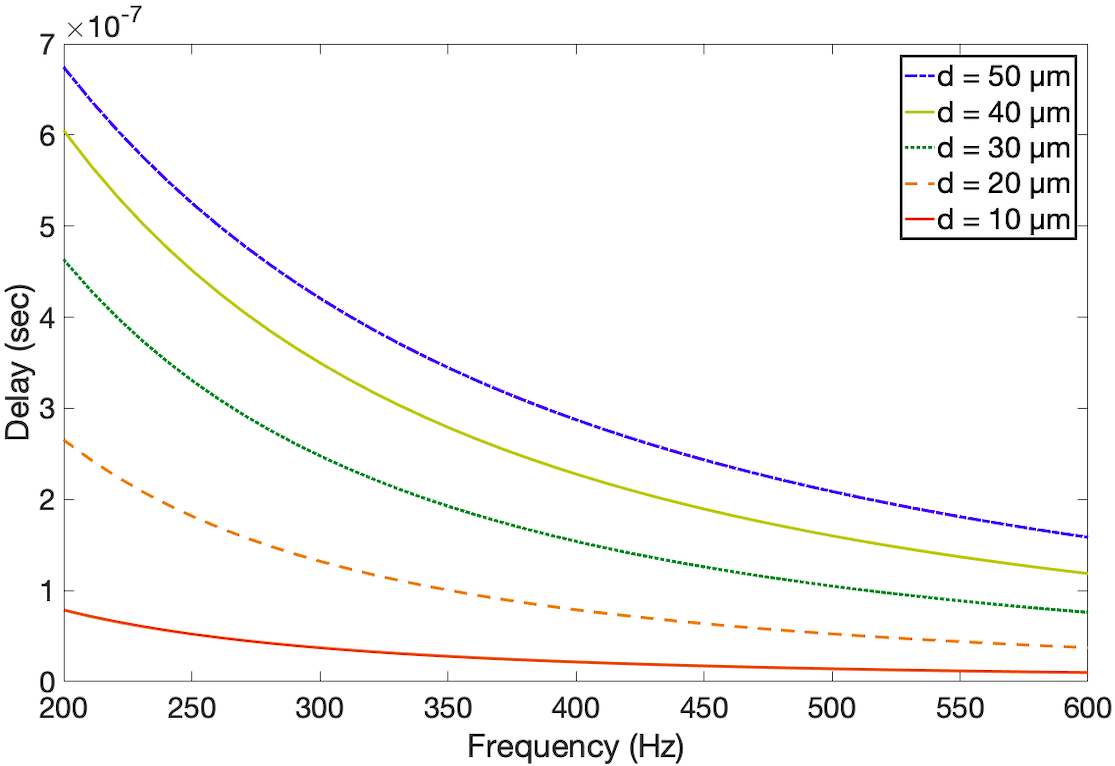}
		\endminipage
		\label{fig:9.1}
		\caption{The electrical characteristics of actin nanowires in terms of attenuation, phase and delay as a function of the frequency and the channel distance.}
	\end{figure*}
	
	\subsection{Channel's Maximum Throughput}     	
	
	The charge capacity of each actin monomer is $\sim$ 4$e$ \cite{tuszynski_ionic_2004}, and by assuming 370 monomers/$\mu s$ of an actin filament, we can deduce that the charge capacity of an actin filament with 1 $\mu m$. long is approximately 1480$e$. By assuming that 1 $e$ represents 1 bit of information, the maximum throughput is calculated by multiplying the total charge capacity of 1 $\mu s$ actin nanowire by the speed of the charges propagation through it and we write \cite{dambri_toward_2019}:
	
	\begin{equation}
	T_M (t) = v(t) \times \psi_{tot}.
	\label{eq:26}
	\end{equation} where $T_M (t)$ represents the nanowire achievable throughput, $v(t)$ is the charge's propagation speed calculated in \cite{hunley_multi-scale_2018} and $\psi_{tot}$ is the total charge capacity of  1 $\mu s$ actin nanowire. Fig. 4 shows the maximum throughput approximation of the proposed nanowire. The decrease of the achievable throughput over time is due to the high attenuation of the actin filaments. However, despite this decrease, and to the best of our knowledge, the throughout is still very high compared to molecular communication results proposed in the literature so far.

	\subsection{Error Probability}
	
	The error probability of the proposed polymer-based channel for wired nano-communication systems can be written as:
	
	\begin{equation}
	P_e = p_0 P_{\epsilon 0} + p_1 P_{\epsilon 1} ,
	\label{eq:27}
	\end{equation} Where $p_0$ = $p_1$= 0.5 are the a priori probabilities, $P_{\epsilon 0}$ is the probability of error for a bit '0' and $P_{\epsilon 1}$ is the probability of error for a bit '1'. The constant assembly and disassembly of actin monomers to construct the actin nanowire generates a noise, which we modeled as a normal distribution. The error is more probable in bit '0' than in bit '1', because the assembly of the nanowire is more frequent than the disassembly. So, the distribution of bit '1' is skewed left compared to the distribution of bit '0', which remains Gaussian and we write \cite{dambri_design_2019}:
	
	\begin{equation}
	\text{\footnotesize $P_{\epsilon 0} =  \frac{1}{\sqrt{2 \pi}} \exp \bigg ({\frac{-x^2}{2}} \bigg )  $},
	\label{eq:28}
	\end{equation} 
	
	\begin{equation}
	\text{\footnotesize $P_{\epsilon 1} =  \frac{1}{\sqrt{2 \pi}}\exp \bigg ({\frac{-(x-A)^2}{2}}\bigg )  \erfc \bigg (\frac{x-A}{\sqrt{2}}\bigg )  $},
	\label{eq:29}
	\end{equation} Where $A$ is the skewness coefficient of the distribution. By substituting (3) and (4) in (2), we write the probability as:
	
	\begin{equation}
	\text{\footnotesize $P_e =  \frac{1}{2\sqrt{2 \pi}} \bigg [ \exp \bigg (\frac{-(x-A)^2}{2}\bigg )  \erfc \bigg (\frac{x-A}{\sqrt{2}} \bigg )  + \exp \bigg ({\frac{-x^2}{2}} \bigg ) \bigg ] $}.
	\label{eq:30}
	\end{equation} 
	
	Since the distribution is negatively skewed, A is also negative. For a skew normal distribution for which the scale factor is 1, the variance is given by $1-\frac{2\delta^2}{\pi}$, where $\delta = \frac{A}{\sqrt {1 + A^2}}$.

	\section{Receiver Design}
	
	From chlorophyll pigments in plants to the neural system in human brains, biological systems always use chemical means to detect and send electrons in the medium. Inspired by muscle fiber contraction and relaxation process, we propose in this paper a receiver design for wired nano-communication networks, which uses SER to detect transmitted electrons through a nanowire. To avoid the absorption of electrons by the receiver's surface, it can be constructed with a hybrid phospholipid/alkanethiol bilayers membrane proposed in \cite{plant_supported_1994}, because of its insulating nature, in order to minimize the loss of electrons at the receiver. The nucleation of monomers that trigger the formation of the actin self-assembly can be tethered at the bio-engineered membrane of the receiver by using a polyethylene glycol (PEG) on one end and an electrode on the other end \cite{hoiles_dynamics_2018}. The insertion of the monomer can be spontaneous, electrochemical or with a proteoliposome insertion as explained in \cite{hoiles_dynamics_2018} with details. When the assembled nanowire reaches the receiver, it binds to one of the  monomers already anchored to the receiver's membrane with electrodes, which creates a passage of electrons through the insulating membrane. As shown in Fig. 1, the designed receiver contains SER that stocks Ca$^{2+}$  ions and a photo-protein that emits a blue light in the presence of Ca$^{2+}$ ions. The emitted light is detected by a photo-detector (gateway).

	\subsection{Electrons Detection}
	
	The Ca$^{2+}$ ion distribution is used in one of the most important biological signaling between living cells, among other functions such as hormone regulation, muscles contraction and neurons excitation \cite{Brini_Intracellular_2013}. The concentration of Ca$^{2+}$ ions inside the cells must be regulated all the time using a very complex system, where SER plays the role of Ca$^{2+}$ ions storage. The smooth endoplasmic reticulum, also called “sarcoplasmic reticulum” in muscle cells, is a tubular structure organelle found in most living cells. The capacity of SER at stocking Ca$^{2+}$ ions is huge because of a buffer called “calsequestrin” that can bind to around 50 Ca$^{2+}$, which decreases the amount of free Ca$^{2+}$ inside SER, and therefore, more calcium can be stored \cite{Katz_Physiology_2010}. When SER is stimulated with an electrical stimulus, the calcium channels open and Ca$^{2+}$ ions get released so fast inside the cells as shown in Fig. 6. The Ca$^{2+}$ concentration is so small inside the cells that a tiny increase in their concentration is detected.

	The SER inside the designed receiver is linked to the nanowire with an electrode so that the transmitted electrons can stimulate the SER, which increases its membrane voltage and open calcium channels. The number of opened calcium channels, and thus, the released concentration of Ca$^{2+}$ is proportional to the intensity of the electrical current stimulating the SER. A small electrical current in the order of picoampere (pA) is sufficient to stimulate SER and release a micromole ($\mu$M) of Ca$^{2+}$\cite{Du_Effects_2009}. When the electrical current stops, calcium channels close and the Ca$^{2+}$ ions will be absorbed and stocked again inside the SER. The short-time presence of Ca$^{2+}$ ions inside the receiver triggers the bioluminescent chemical reaction, which emits a blue light.
	
	\subsection{Light Emission}
	
	Photo-proteins are priceless biochemical tools for a variety of fields including drug discovery, protein dynamic studies and gene expression analysis. \textit{Aequorin} is a very important photo-protein for biological studies because it helps researchers measure and study the Ca$^{2+}$ distribution in vivo. Upon binding to Ca$^{2+}$ ions, the oxidation of \textit{Aequorin} molecule is triggered, resulting in the emission of a bioluminescent blue light (470 nm). There are other photo-proteins in nature with different wavelengths emissions such as \textit{Luciferin} (530 nm) and some chromophores (630 nm). However, unlike other bioluminescent reactions that involve the oxidation of an organic substrate such as \textit{Luciferin} and chromophores, adding molecular oxygen is not required in Ca$^{2+}$-dependent light emissions, because \textit{Aequorin} protein has oxygen bound to it \cite{Weeks_Signal_2013}. Other advantages of using \textit{Aequorin} are that it does not involve any diffusible organic factor, no direct participation of enzymes and that it can be recycled after use \cite{Shimomura_Properties_1969}. \textit{Aequorin}  is extracted from the jellyfish \textit{Aequorea Victoria} that lives in North America and the Pacific Ocean \cite{shimomura_short_1995}, and then purified in distilled water. This photo-protein is very sensitive to changes in the concentration of free Ca$^{2+}$, which explains its extensive use as an indicator of Ca$^{2+}$ ions in biological studies. In the presence of a very low concentration of Ca$^{2+}$, the light intensity emitted by \textit{Aequorin}  will be independent from Ca$^{2+}$ concentration, the bioluminescent reaction becomes Ca$^{2+}$-dependent only when the concentration exceeds 10$^{-7} $ M. 
	
	The designed receiver uses the short-time presence of the released Ca$^{2+}$ ions to trigger the oxidation of Aequorin, which emits a photon for each 3 Ca$^{2+}$ ions bound, as shown in Fig. 7. This mechanism of emitting-light can trigger the release of over 70 kcal of energy as a visible radiation through a single transition \cite{Shimomura_Properties_1969}. In the next section, we model the electron detector as an RC circuit, and we calculate the radiant energy emitted from the bioluminescent reaction at each symbol interval.
	
	\section{Receiver Model}
	
	Cell membranes are biological structures that surround cells and separate their interior from the external environment for the protection and control of ion exchanges. The membrane is constructed with phospholipid molecules, where the lipid end is hydrophobic and the phosphate end is hydrophilic. When lipid ends get together to form a double-layered sheet and close the sphere, they create a spherical surface that perfectly separates two volumes of liquid. However, a pure phospholipid bilayer is an excellent insulator which does not allow any ion exchange, thus, other pores penetrating the membrane are necessary, allowing the ions to circulate inside and outside the cell. These pores are called “ion channels” and by controlling their opening and closing, the cell controls the ions flow. Cell membranes have specific channels for each ion type, and closing them creates a difference between the ion's concentration inside and outside the cell, which set up a specific potential for each ion type.
		
	\begin{figure}
		\centering
		\includegraphics[width=\linewidth]{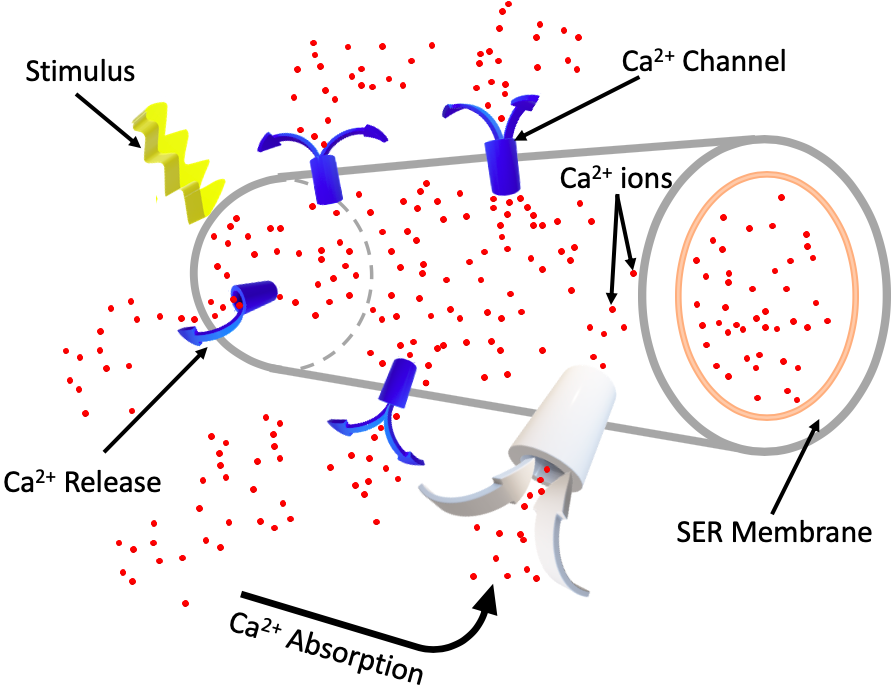}
		\caption{Ca$^{2+}$ ions release by a Smooth Endoplasmic Reticulum (SER).}
		\label{fig:2}
	\end{figure}
	
	\begin{figure}
		\centering
		\includegraphics[width=\linewidth]{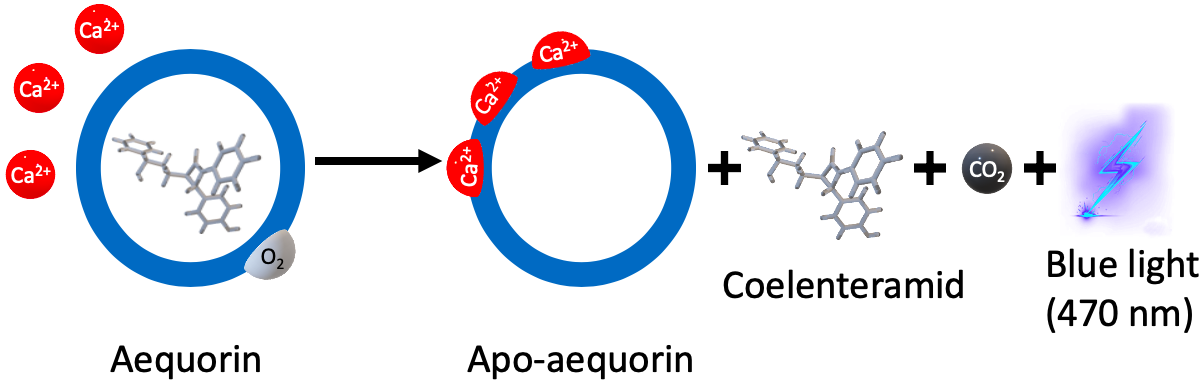}
		\caption{Aequorin bioluminescent reaction that generates blue light in the presence of Ca$^{2+}$ ions.}
		\label{fig:3}
	\end{figure}

	Let's consider $V_i$ the reversal potential of ion species \textit{i}, which is the value of the membrane potential for which the flux of ion species \textit{i} is zero, and let's consider $R_i$ the channel resistance, which is simply the inverse of channel's conductance. According to Ohm's law, the ions flow across the channel is proportional to the reversal potential, and the proportionality factor is the channel conductance $g_i$ and we can write:
	
	\begin{equation}
	I_i= \frac{V_i}{R_i} = g_i V_i,
	\label{eq:1}
	\end{equation} where $I_i$ is the current of ion species \textit{i} that flows across the membrane. The current flows until reaches an equilibrium called the resting potential $V_0$, which can be calculated as:

	\begin{equation}
	V_0= \frac{\sum_i g_i V_i}{\sum_i g_i},
	\label{eq:2}
	\end{equation}
	
	In our case, the membrane has only specific channels for Ca$^{2+}$ ions, therefore, $g_i$ becomes $g_{Ca}$ and $V_i$ becomes $V_{Ca}$. The reversal potential $V_{Ca}$ is calculated by using Nernst equation as:
	
	\begin{equation}
	V_{Ca}= \frac{kT}{ze} \ln{\frac{P_{out}}{P_{in}}},
	\label{eq:3}
	\end{equation} where $k$ is the Boltzmann constant, $e$ is the electron charge, $T$ is the temperature in Kelvin, $z$ is the valence of the Ca$^{2+}$ ion, $P_{out}$ and $P_{in}$ are the probabilities of finding a Ca$^{2+}$ ion outside or inside the SER respectively. When the membrane is at rest, $V_{Ca}$ equals $V_0$, however, when an external potential is applied, $V_{Ca}$ increases which opens the Ca$^{2+}$ channels and discharges SER from Ca$^{2+}$ ions. When the excitation is passed, special pores called pumps charges SER again with Ca$^{2+}$ ions. The charge and discharge of SER can be modeled as an equivalent capacitance and the inverse of the channel conductance can be modeled as an equivalent resistance in series with a voltage source. Therefore, the SER and its membrane can be modeled as an equivalent RC circuit as shown in Fig. 8.
	
	\subsection{The Capacitance}
	
	The SER membrane separates two conductive liquids that contain free ions, so we have two conductors separated by an insulator. The potential difference across the membrane that separates a charge of a Ca$^{2+}$ ion $Q_{Ca}$ defines a capacitance $C_{Ca}$ that can be written as:
	
	\begin{equation}
	C_{Ca}= \frac{Q_{Ca}}{\mid \Delta V_{Ca} \mid },
	\label{eq:4}
	\end{equation}
	
	Before calculating the potential difference $\Delta V_{Ca}$ by using the Gauss's law, we define the permittivity $\varepsilon$ of the membrane as $\varepsilon=\varepsilon_0 \varepsilon_r$, where $\varepsilon_r$ is the relative permittivity. SER in muscle cells is composed of tubule networks called cisternae, which they have a diameter $r_{SER}=50 nm$. The SER tubules extend throughout muscle fiber filaments, so we assume that the length of SER tubules used in our designed receiver is $l =1\mu m$. By assuming that the enclosed charge within SER's volume is $Q_{Ca}$, then from Gauss's law we can write the electric field at a distance $r$ as:
	
	\begin{equation}
	\vec E_{Ca}= \frac{Q_{Ca}}{\varepsilon} \bigg (\frac {1}{2\pi lr} \bigg ) \hat r.
	\label{eq:5}
	\end{equation} where $\hat r$ is the radial vector from the Ca$^{2+}$ charge at the origin of SER tubules to the surface of the membrane. The potential difference between the inside and the outside of SER membrane is therefore written as:
	
	\begin{equation}
	\text{\footnotesize $\Delta V_{Ca}= -\int_{r_{SER}}^{r_{SER}+\delta} E_{Ca}.dr=-\frac {Q_{Ca}}{2\pi lr} \ln{\bigg (1+ \frac {\delta}{r_{SER}} \bigg )} $},
	\label{eq:6}
	\end{equation} where $\delta$ is the thickness of the membrane. By substituting eq. 6 in eq. 4 we find:
	
	\begin{equation}
	C_{Ca}= \frac{2\pi lr}{\ln{\big (1+ \frac {\delta}{r_{SER}}\big )}}.
	\label{eq:7}
	\end{equation}
	
	The membrane of SER does not need a high voltage to separate charges because it is only two molecules thick ($\delta=6\times10^{-9} m$), thus, we expect the membrane capacitance to be quite high per unit area. With the parameters given above, we estimate that the capacitance is $C_{Ca}\simeq 5\times10^{-6}pF/\mu m^2$. Unlike the conductance, the capacitance of biological membranes is not influenced by the complexities of biological systems, which makes it constant.

	\begin{figure}
		\centering
		\includegraphics[width=\linewidth]{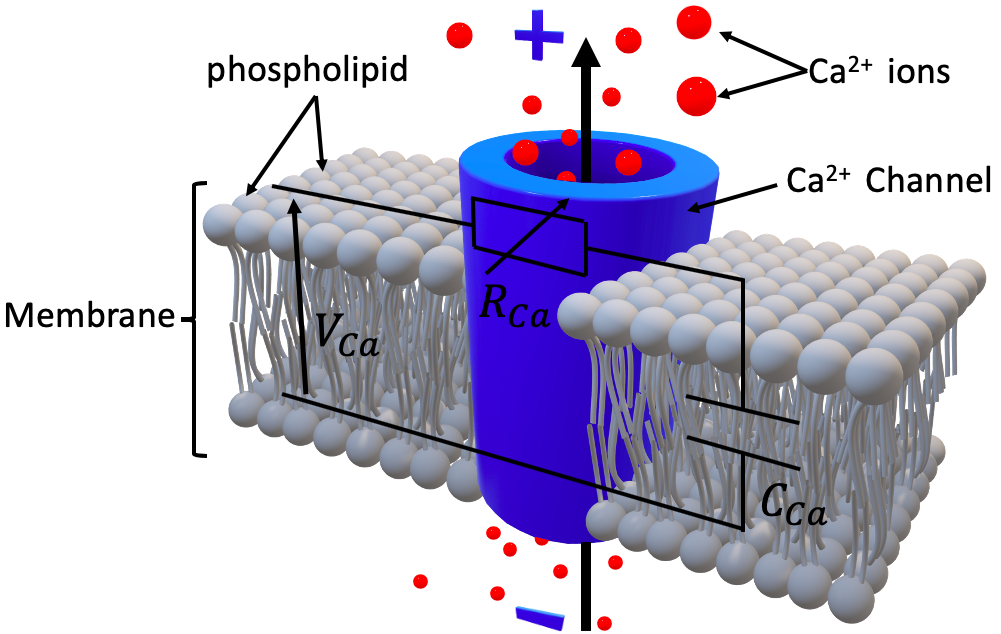}
		\caption{The proposed equivalent RC circuit of SER's membrane.}
		\label{fig:4}
	\end{figure}

	\subsection{The Resistance}
	
	The pure phospholipid bilayer constructing the membrane is an excellent electrical insulator with a very low conductance, but the mosaic proteins that span the surface of the membrane act as channels for ions and increase the conductance. The Ca$^{2+}$ current flowing through SER membrane depends on the number of open channels by the applied external voltage. Depending on the type and the age of SER, the number of channels varies between 3.8 and 24.3 per $\mu m^2$ area \cite{Orchard_T-tubules_2007}. By taking into account the infinitesimal area of channels, we consider the SER membrane as an infinite charged linear line and by using Gauss's law we can write the Ca$^{2+}$ current $I_{Ca}$ of one channel as:
	
	\begin{equation}
	I_{Ca}= \sigma \int E_{Ca}.da=\sigma \frac {\lambda l}{\varepsilon},
	\label{eq:8}
	\end{equation} where $\sigma$ is the conductivity of one Ca$^{2+}$ channel and $\lambda=Q_{Ca}/l$. From eq. 1 we can calculate the resistance of Ca$^{2+}$ channels $R_{Ca}$:
	
	\begin{equation}
	R_{Ca}= \frac{V_{Ca}}{I_{Ca}},
	\label{eq:9}
	\end{equation} and by substituting eq. 6 and eq. 8 in eq. 9 we find: 
	
	\begin{equation}
	R_{Ca}= \frac{\ln{\big (1+ \frac {\delta}{r_{SER}}\big )}}{2\pi n\sigma l}.
	\label{eq:10}
	\end{equation} where $n$ is the number of Ca$^{2+}$ channels. $R_{Ca}$ is the sum of n variable resistances linked in parallel in our equivalent circuit. The conductivity of a Ca$^{2+}$ channel is estimated to be $\sigma=0.5 pS$ \cite{Krishtal_Conductance_1981}, and thus, the mean value of the variable resistance in eq. 9 can be approximated to $R_{Ca}\simeq 6.11M\Omega$. The membrane resistance is highly variable because it is so thin that a very small change in the voltage can create a strong electric field within it.
	
	\subsection{Radiant Energy}
	
	The presence of Ca$^{2+}$ ions near the photo-protein Aequorin triggers a bioluminescent reaction that emits a blue light. We assume that Ca$^{2+}$ ions and \textit{Aequorin} molecules have homogenous distributions inside the receiver and that for each 3 Ca$^{2+}$ ions released outside SER, there is an \textit{Aequorin} molecule that binds to them and emits one photon. Radiant energy is an electromagnetic energy carried by a stream of photons, thus, the sum of photons emitted by the bioluminescent reaction represents the radiant energy of the designed receiver. With the assumptions we made above, we calculate the radiant energy of the designed receiver for each symbol interval as follows:
	
	\begin{equation}
	Q_e= \frac{1}{3} \frac{hc}{\lambda} \int_{0}^{T} p.c(t) dt,
	\label{eq:11}
	\end{equation} where $h$ is the Planck constant, $c$ is the speed of light, $\lambda$ is the photon wavelength. $T$ is the symbol time, and $p$ is the probability for each \textit{Aequorin} molecule to emit a photon. In this paper, the concentration of the released Ca$^{2+}$ ions for each symbol interval $c(t)$ is obtained numerically by using Simulink in MATLAB, which represents the output of the capacitor in the proposed equivalent circuit. An analytical expression for the released Ca$^{2+}$ ion concentration will be derived in our future work. 
	
	\section{Modulation Techniques}
	
	The process of encoding information by varying one or more properties of the carrier signal is called modulation. A variety of modulation techniques have been proposed for molecular communications networks by changing the concentration of molecules, molecules  type or the time of the molecules release \cite{farsad_comprehensive_2016}. 
	
	The first proposed method is the Concentration Shift Keying (CSK) where a bit is decoded at the receiver as "1" if the molecules concentration reaches a predefined threshold at a symbol interval, otherwise it is a "0" \cite{kuran_interference_2012}. Molecular Shift Keying (MoSK) is another modulation method proposed by the same authors that uses $2^n$ molecule types to transmit $n$ bits of information \cite{kuran_interference_2012}. In \cite{Garralda_Diffusion_2011}, the authors proposed Pulse Position Modulation (PPM) that encodes the information by changing the time of molecules release. They divided the symbol interval into two equal halves, and the receiver decodes the bit as "1" if the pulse is in the first half, and as "0" if the pulse is in the second half. However, the modulation techniques proposed for molecular communication cannot be used for wired nano-communication networks, which they use electrons as carriers of information instead of molecules. Therefore in this paper, we propose a new modulation technique for wired nano-communication networks that encodes the information by changing the intensity of the bioluminescent light, the time between two consecutive light emissions or a combination between the intensity and the time changes. We call these methods Bioluminescence Intensity Shift Keying (BISK), Bioluminescence Time Shif Keying (BTSK) and Bioluminescence Asynchronous Shift Keying (BASK) consecutively.
	
	\subsection{Bioluminescence Intensity Shift Keying (BISK)}
	
	The total emitted light capacity of the designed receiver is proportional to the released Ca$^{2+}$ concentration and the amount of recycled \textit{Aequorin} for each symbol interval, which is proportional to the pulsed electrical intensity transmitted through the nanowire. Therefore, we can modulate the transmitted information by varying the intensity of the bioluminescent light emitted at the receiver. The change in the intensity is detected at the gateway, where an appropriate threshold is used to correctly extract the information sent through the nanowire. A bit is decoded as "1" if the blue light intensity $Q_e$ in eq. 16 exceeds a threshold $\xi$, and as "0" otherwise.
	
	The proposed BISK modulation technique has the advantage of reducing the Inter-Symbol Interference (ISI) which is caused by the remaining molecules from a previous symbol. Even if some  Ca$^{2+}$ ions remain in the medium causing the emission of some photons, the intensity of light is still low, and we just need a photo-detector with high resolution at the gateway to eliminate the influence of ISI on the receiver. 
	
	\subsection{Bioluminescence Time Shift Keying (BTSK)}
	
	As in PPM modulation technique proposed in \cite{Garralda_Diffusion_2011}, we can divide our symbol interval into two equal halves. The bit is decoded as "1" if the bioluminescent light is detected in the first half, and as "0" if it is detected in the second half. This method can be a little bit slower than BISK, because detecting small changes in the light intensity is much faster than waiting a hole symbol to decode one bit. However, BTSK can be more efficient in terms of bit error rate, especially when the majority of bits have the same value "1" or "0". 
	
	\subsection{Bioluminescence Asynchronous Shift Keying (BASK)}
	
	Another technique that we propose is a hybrid modulation scheme based on a combination between BISK and BTSK. This modulation technique is asynchronous, and it can achieve a higher information rate than the intensity-based and time-based approaches separately. By using $2^k-1$ threshold levels, BASK can use $k$ bits with $2^k$ different values to represent one symbol, which increases the number of bits per symbol and thus increases the information rate. Then $k=2$ when we encode one bit in the intensity chance and another bit in the time change. Therefore, by using the classical modulation naming convention based on the number of bits per symbol, BASK modulation technique can be called Quadruple BASK (QBASK) when $k=2$.
	
	In this paper  we focus on Bioluminescence Intensity Shift Keying (BISK) modulation technique. The other two proposed modulation techniques will be covered in our future work.
	
	\subsection{Bits Decoding}
	
	Since the emission of a bioluminescent light depends on photo-proteins reacting with the Ca$^{2+}$ ions released inside the receiver which exhibit Brownian motion, a single photon has a certain probability $P_{em}$ of being emitted and detected at the gateway. This probability $p=P_{em}(r, t_b)$ defined in eq. 16 depends on the number of the recycled photo-protein per symbol interval $r$ and the bit duration $t_b$. Because of the huge number of Ca$^{2+}$ ions released by SER in each bit duration, we can safely assume that each recycled photo-protein can react with three Ca$^{2+}$ ions at the same time. Thus, the reaction explained in Fig. 7 is considered as a single event instead of three separate events, whether the photo-protein is activated to emit a photon or not. We also assume that both Ca$^{2+}$ ions and photo-proteins are uniformly distributed inside the receiver and that the bioluminescent reaction happens instantaneously. Based on the assumptions discussed above, if $c$ ions are released in a symbol duration, the number of photons emitted by reacting with these $c$ ions is a random variable following a binomial distribution:
	
	\begin{equation}
	C \sim Binomial (c, P_{em}(r, t_b))
	\label{eq:17}
	\end{equation} 
	
	The Binomial distribution can be approximated to a normal distribution by Central Theorem Limit (CLT) \cite{wilkinson_stochastic_2011}. When $cp$ is large enough and $p$ is not closer to zero or one, the distribution of the photons emission variable $C$ can be written as:
	
	\begin{equation}
	C_n \sim \mathcal{N}(cp, cp(1-p))
	\label{eq:18}
	\end{equation} 
	
	We assume that the noise inside the designed receiver is Additive Gaussian White Noise (AWGN). Therefore, the noise is also a random variable that follows a normal distribution:
	
	\begin{equation}
	N \sim \mathcal{N}(0, \sigma^2)
	\label{eq:18}
	\end{equation} and thus, the probability of detecting the bioluminescent light emitted at the receiver can be found by adding the two normal distributions. The probability of detecting a previous bit is neglected in this paper by assuming that the photo-detector used at the gateway has a high resolution. 
	
	By using the proposed BISK modulation technique, the gateway decodes the detected bioluminescent light by comparing the sum of the two normal distributions with a predefined threshold value $\xi$. If the intended bit is "0", then the probability of an incorrect decoding for the BISK modulation technique are calculated as follows:
	
	\begin{equation}
	P_{r} = \frac{1}{3} \frac{hc}{\lambda} P(C_n+N \geqslant \xi)
	\label{eq:19}
	\end{equation} and if the intended bit is "1":
	
	\begin{equation}
	P_{r} = \frac{1}{3} \frac{hc}{\lambda} P(C_n+N < \xi)
	\label{eq:20}
	\end{equation} 
	
		\begin{figure}
		\centering
		\includegraphics[width=\linewidth]{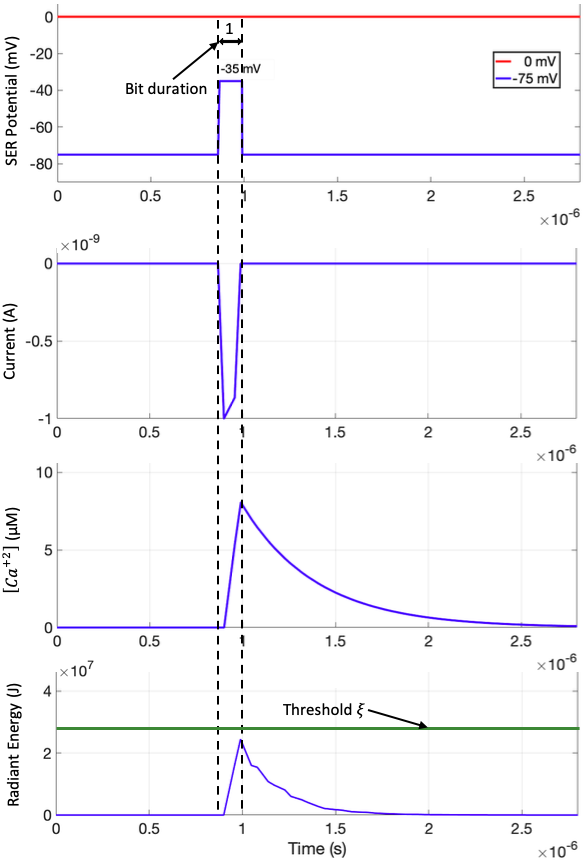}
		\caption{Bioluminescence Intensity Shift Keying (BISK) modulation technique.}
		\label{fig:5}
		\end{figure}
	
	\begin{figure*}[!htb]
		\minipage{0.32\textwidth}
		\includegraphics[width=\columnwidth]{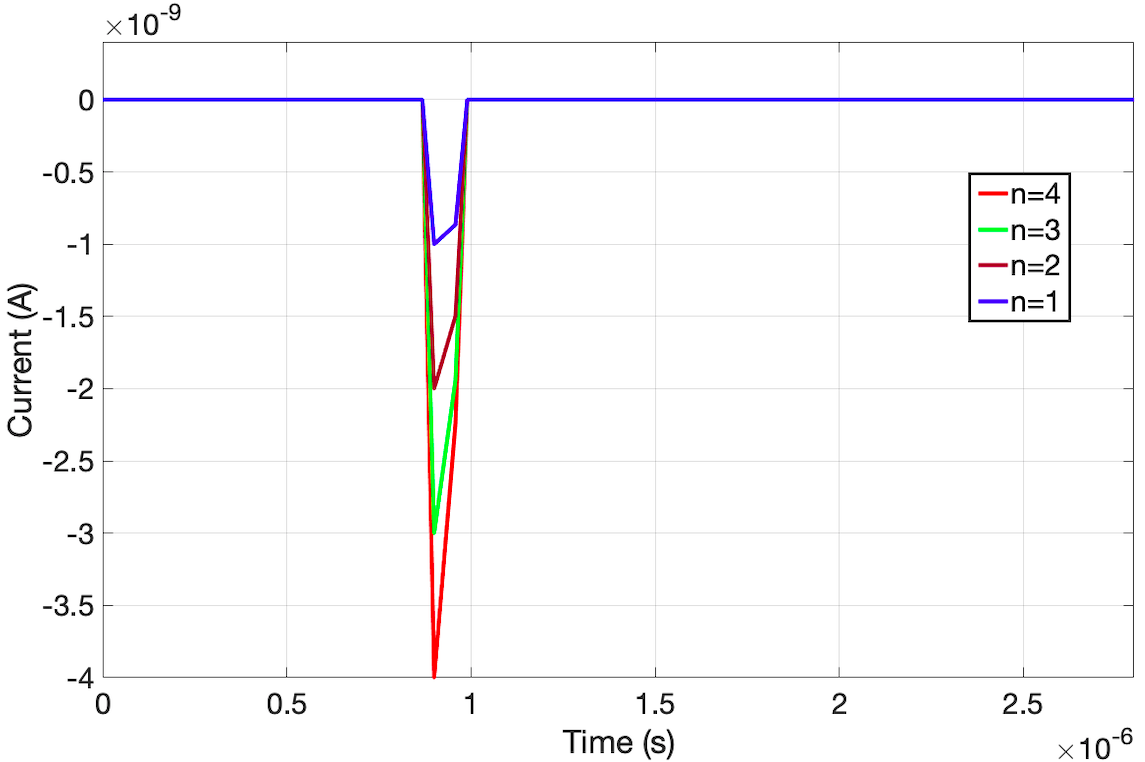}
		\caption{The current necessary to open the Ca2+ channels.}
		\label{fig:6a}
		\endminipage\hfill
		\minipage{0.32\textwidth}
		\includegraphics[width=\linewidth]{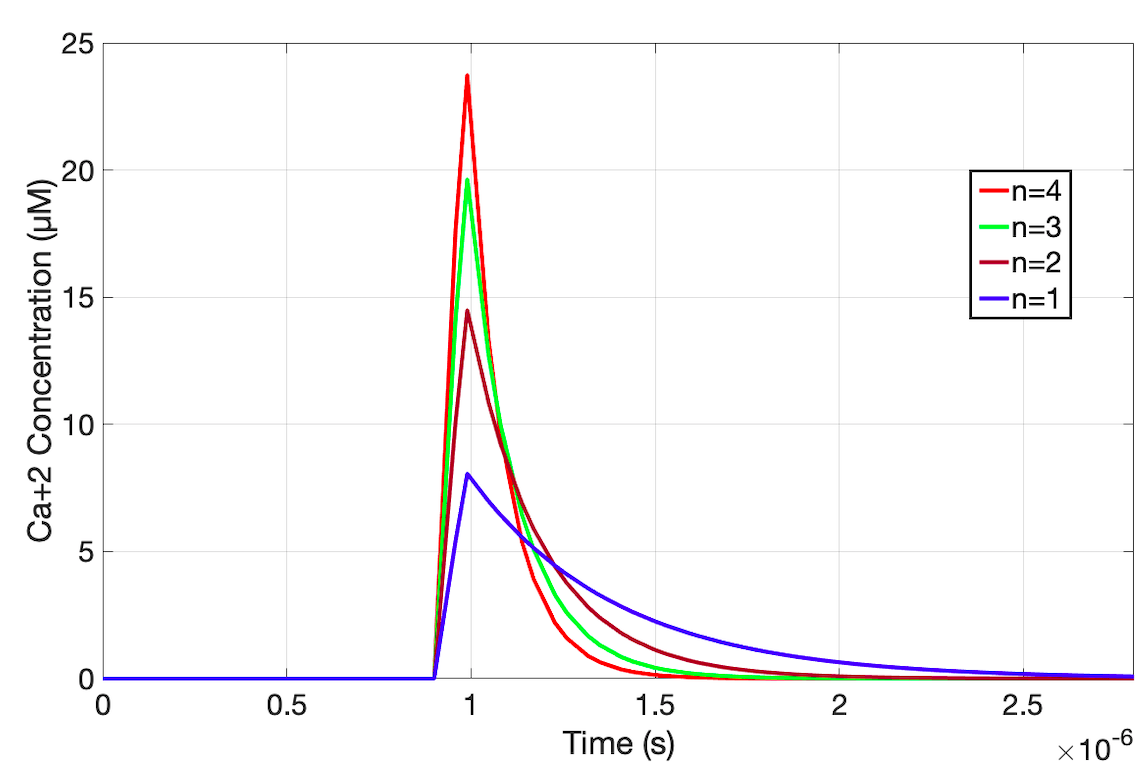}
		\caption{The released Ca2+ ion concentration for each number of open channels.}
		\label{fig:6b}
		\endminipage\hfill
		\minipage{0.32\textwidth}
		\includegraphics[width=\linewidth]{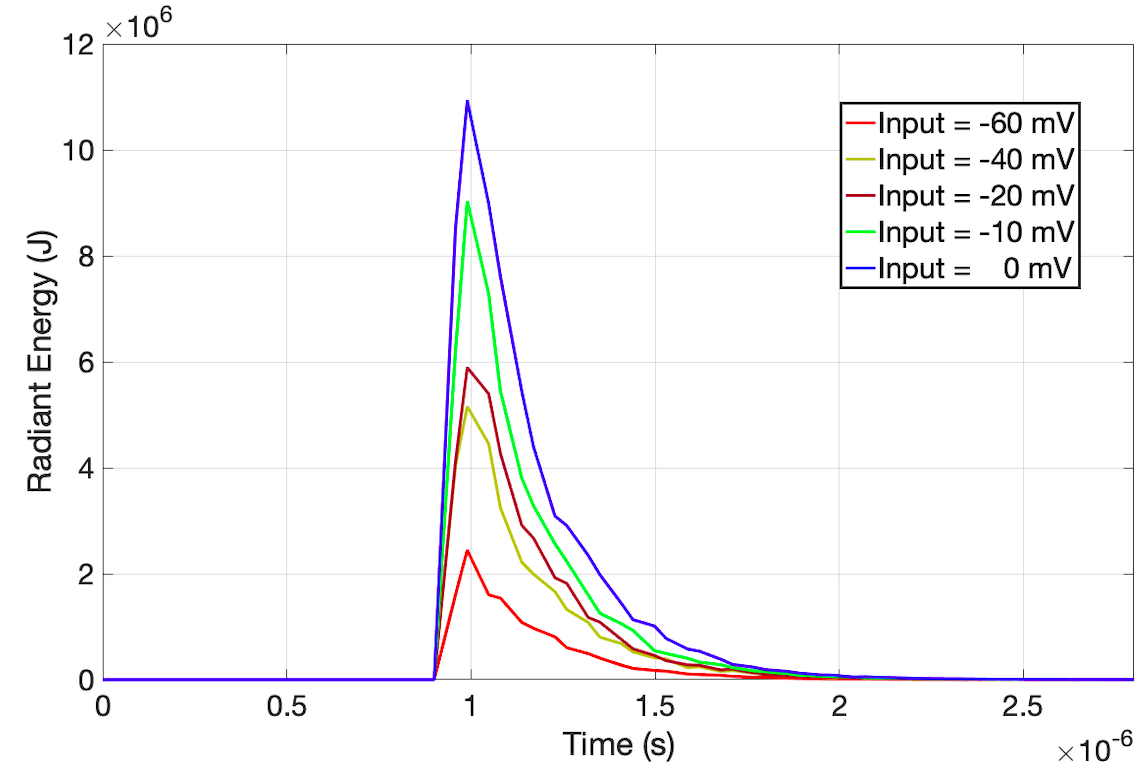}
		\caption{Radiant energy emitted by the bioluminescent reaction for different inputs.}
		\label{fig:6c}
		\endminipage\hfill
	\end{figure*}
	
	\section{Performance Evaluation and Numerical Results}
	
	To evaluate the performance of the designed receiver, we used Simulink in MATLAB to simulate the proposed equivalent circuit. The parameters we used in this study are mentioned in the section above. The output at the capacitance represents the released Ca$^{2+}$ ion concentration by SER for each bit interval. As shown in Fig. 9, the current sent through the nanowire excites SER’s membrane and increases its resting potential (-75 mV) calculated in eq. 2 by 35 mV, which allows the opening of Ca$^{2+}$ channels. The output of the capacitance increases rapidly to reach more than 7 $\mu M$ of Ca$^{2+}$ ions, then when the excitation stops, SER starts absorbing the released Ca$^{2+}$ ions and the output decreases slowly until it vanishes from the medium.
	
	\subsection{Number of Ca$^{2+}$ Channels}
	
	In this study, we used the minimum possible channels number per $\mu m^2$ area in SER membrane, which is $n=4$ channels. These 4 channels represent 4 resistances linked in parallel in our proposed equivalent circuit. We simulated 4 cases where only one channel is opened, two, three and four channels opened together and we plotted the current necessary for each case along with the output of the equivalent circuit as a function with time.  
	
	In Fig. 10, we can see that the number of open channels is proportional to the intensity of the transmitted current through the nanowire. We can also notice that only an intensity of 4 pA is needed to open the maximum number of channels we used in this study. This is due to the fact that the membrane's thickness is so small (6 nm) that a tiny excitation can change the voltage at the channels and open them. The sensitivity of SER’s membrane to the electric current is one of the reasons why we use it to detect electrons in our designed receiver for a wired nano-communication.

	Fig. 11 shows the output results of the proposed equivalent circuit in the 4 studied cases. We can notice that with 4 opened channels, SER can release more than 20 $\mu M$ of Ca$^{2+}$ ions. We can also notice that in the case where 4 channels are open, the signal decreased rapidly (1.5 $\mu s$) compared to the case where only one channel is open (3 $\mu s$).  With 4 opened channels, the SER needs half the time that one channel needs to absorb the released Ca$^{2+}$ ions after the excitation is stopped, which deceases the intersymbol interference between previous and next bits. The released Ca$^{2+}$ concentration by SER is proportional to the intensity of the current sent through the nanowire, which is used to demodulate the transmitted signal.
	
	\subsection{Radiant Energy}
	
	We used the simulated output of the equivalent circuit which represents the concentration of the released Ca$^{2+}$ ions for each bit interval $c(t)$ to calculate the bioluminescent radiant energy by using eq. 11. We used different voltage intensities at the input to show the proportionality between the transmitted current through the nanowire and the light intensity emitted by the receiver. Fig. 12 shows that the intensity of the blue light emitted by the receiver can reach more than 10 MJ, which is enough to be detected by the gateway. It also shows that the intensity of light decreases rapidly when the medium is emptied of Ca$^{2+}$ ions, because without the presence of Ca$^{2+}$ ions, the bioluminescent light cannot be emitted, which decreases the intersymbol interference.

	\begin{figure}
		\centering
		\includegraphics[width=\linewidth]{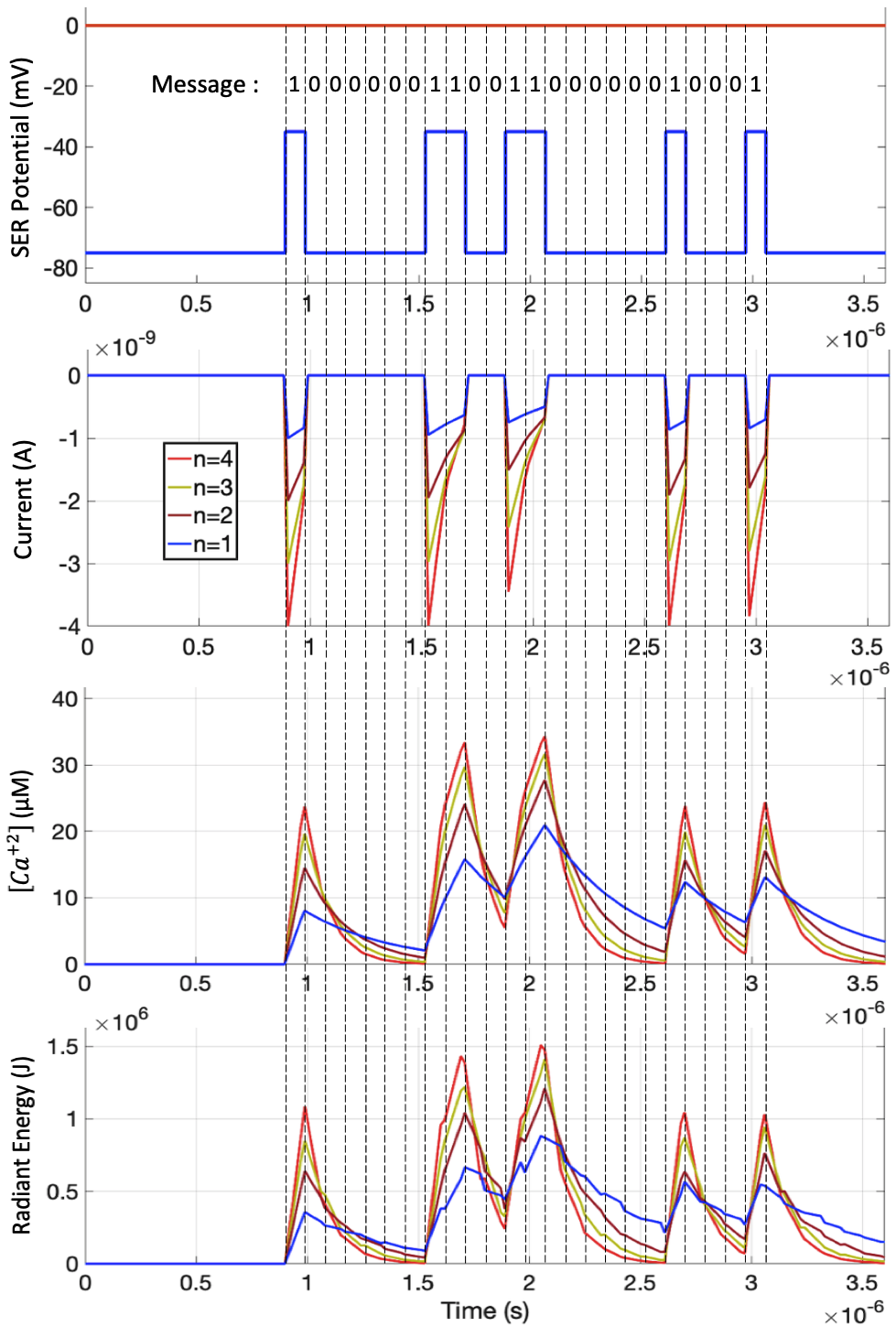}
		\caption{The response of the designed receiver to a random binary message.}
		\label{fig:9}
	\end{figure}
	
	Fig. 13 summarizes the performance of the designed receiver by following the electrons detection and the light emission processes. The figure shows a random binary message that is detected by SER, where the bits "1" open the channels and release Ca$^{2+}$ ions, which they combine with \textit{Aequorin} and emit a blue light. Bit "0" keep the channels closed, SER absorbs the Ca$^{2+}$ ions from the medium and the light emission stops. The results of this study show that light emission allows the designed receiver to play the role of a relay between the nanomachines and the gateway. This ability can be used as a link between the nanoscale and macroscale in the in-body technologies for medical applications.

	\begin{figure}
		\centering
		\includegraphics[width=\linewidth]{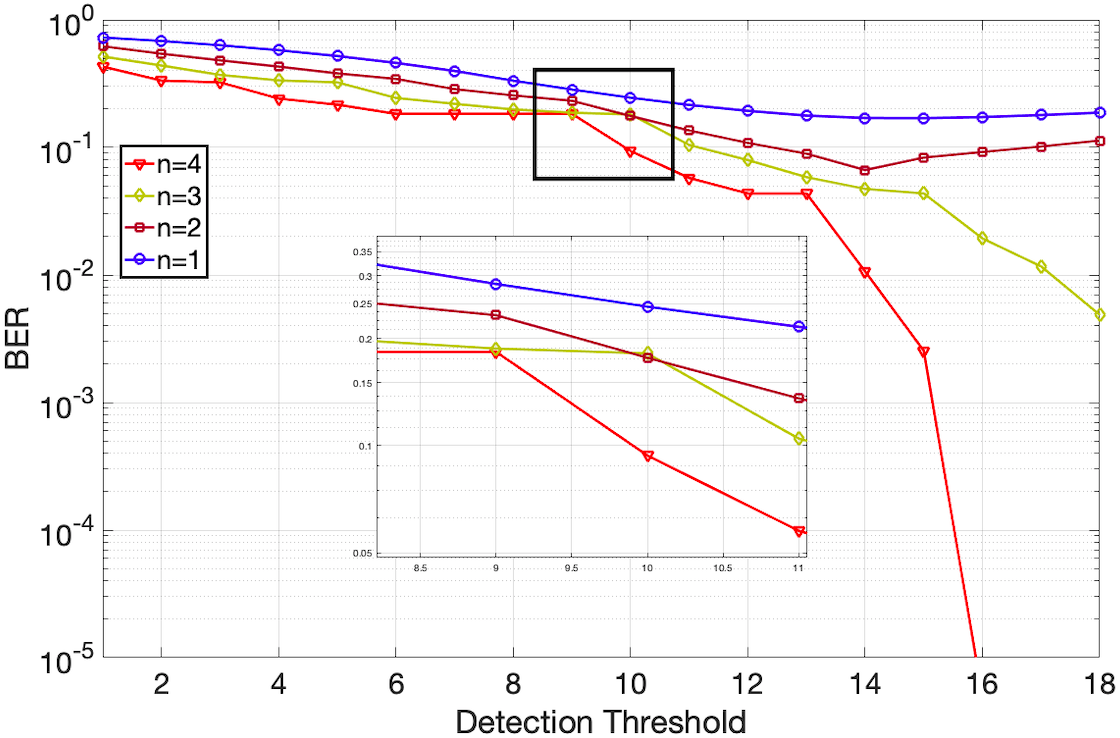}
		\caption{Bit error rate of the designed receiver as a function of the detection threshold.}
		\label{fig:10}
	\end{figure}
	
	\subsection{Bit Error Rate}
	
	In order to evaluate the performance of the designed receiver, we calculate the Bit Error Rate (BER), which is the bit errors number divided by the total transmitted bits during the studied time interval. Fig. 14 shows the results of the calculated BER as function of the detection threshold for the 4 studied cases. We notice that the case with 4 open channels has the best results with BER of $10^{-5}$ for a threshold 16, because 4 channels absorb the released Ca$^{2+}$ ions faster, which deceases the intersymbol interference as we explained above. The fewer opened channels at the membrane of SER the more bit errors will be received, where 3 opened channels can reach BER of $10^{-2}$ for a threshold 17. The worst case is where only one channel is opened with 0.8, which is explained by the fact that one single channel will not be able to release and absorb Ca$^{2+}$ ions rabidly, which increases intersymbol interferences, and thus, increases the BER.
	
	\section{Conclusion}
	
	Using the ability of certain polymers to self-assemble in order to build conductive nanowires between nanomachines is a new method proposed for establishing wired nano-communication systems. Despite the very high achievable throughput of wired nano-communication systems due to the use of electrons as information carriers, the detection of electrons on the nanometric scale is very challenging. This paper proposes a bio-inspired receiver design that uses Smooth Endoplasmic Reticulum (SER) to detect the transmitted electrons by measuring the released Ca$^{2+}$ ions concentration with the photo-protein \textit{Aequorin} that emits a blue light in the presence of Ca$^{2+}$ ions. To better design the receiver, we simulate the construction of the nanowire, present its electrical characteristics and calculate its maximum throughput. We modeled the dynamic of SER with an equivalent circuit, and we derived the analytical expression of their components. We calculated the radiant energy emitted in each symbol interval, and we simulated the detection and light emission processes of a random binary message. We also proposed modulation techniques that guaranty an effective decoding of the information and we calculated the BER of the designed receiver to evaluate its performance. The results of this study showed that the designed receiver is efficient for wired nano-communication networks and that it can also play the role of a relay for the nearest gateway.
	
	\bibliographystyle{IEEEtran}
	\bibliography{ref}
	
\end{document}